\documentclass{svproc}
\usepackage{url}

\usepackage{graphicx}
\usepackage{float}
\usepackage{amsmath}
\usepackage{multirow}
\usepackage{tabularx}
\usepackage{placeins}
\usepackage{flafter}
\usepackage{booktabs}
\usepackage{array}
\newcolumntype{L}[1]{>{\raggedright\arraybackslash}p{#1}}
\usepackage{longtable} % if table spans pages
\usepackage[colorlinks=true,citecolor=blue,linkcolor=blue,urlcolor=blue]{hyperref}
\usepackage{cite}
\usepackage{wasysym}

\usepackage{xcolor}
\usepackage{pgf}
\usepackage{multirow}
\usepackage{xfp}   

\newcommand{\before}[1]{\textcolor{blue}{#1}}
\newcommand{\after}[1]{\textcolor{red}{#1}}

\definecolor{pctA}{HTML}{EEF9EE} % 0–2
\definecolor{pctB}{HTML}{DCF3DC} % 2–4
\definecolor{pctC}{HTML}{C8ECC8} % 4–6
\definecolor{pctD}{HTML}{B2E3B2} % 6–8
\definecolor{pctE}{HTML}{99D999} % 8–10
\definecolor{pctF}{HTML}{7ECF7E} % 10–12
\definecolor{pctG}{HTML}{63C563} % 12–14+

\newcommand{\pctbox}[1]{%
  \begingroup
  \pgfmathtruncatemacro{\bin}{min(6, int(#1/2))}%
  \def\pctcol{pctA}%
  \ifcase\bin
    \def\pctcol{pctA}%
  \or\def\pctcol{pctB}%
  \or\def\pctcol{pctC}%
  \or\def\pctcol{pctD}%
  \or\def\pctcol{pctE}%
  \or\def\pctcol{pctF}%
  \or\def\pctcol{pctG}%
  \fi
  \text{\colorbox{\pctcol}{\scriptsize(+#1\%)}}%
  \endgroup
}

\definecolor{gainLow}{HTML}{EEF9EE}        % 0–10
\definecolor{gainMild}{HTML}{D6F2D6}       % 11–20
\definecolor{gainModerate}{HTML}{BDEABD}   % 21–30
\definecolor{gainHigh}{HTML}{97D997}       % 31–40
\definecolor{gainExtreme}{HTML}{63C563}    % 41–50+

\newcommand{\pctboxL}[1]{%
  \begingroup
  \edef\p{\fpeval{#1}}%
  \def\pctcol{gainLow}%
  \ifnum\fpeval{(\p>10)?1:0}=1 \def\pctcol{gainMild}\fi
  \ifnum\fpeval{(\p>20)?1:0}=1 \def\pctcol{gainModerate}\fi
  \ifnum\fpeval{(\p>30)?1:0}=1 \def\pctcol{gainHigh}\fi
  \ifnum\fpeval{(\p>40)?1:0}=1 \def\pctcol{gainExtreme}\fi
  \text{\colorbox{\pctcol}{\scriptsize(+#1\%)}}%
  \endgroup
}

\begin{document}
\mainmatter              % start of a contribution
\title{Strengthening Human-Centric Chain-of-Thought Reasoning Integrity in LLMs via a Structured Prompt Framework}

%Enhancing the Performance of Human-Centric Chain-of-Thought Reasoning in Large Language Models By Prompt Engineering

%\titlerunning{Contribution Title}  % abbreviated title (for running head)
%                                     also used for the TOC unless
%                                     \toctitle is used
\author{Jiling Zhou \inst{}$^{*}$  \and Aisvarya Adeseye\inst{} \and Seppo Virtanen \inst{} \and \\Antti Hakkala \inst{} \and Jouni Isoaho \inst{}}
\authorrunning{Jiling Zhou et al.} % abbreviated author list (for running head)
\titlerunning{Strengthening Human-Centric Chain-of-Thought Reasoning Integrity} 

\institute{Department of Computing, University of Turku, 20014 Turku, Finland\\
\email{$^{*}$jizhou@utu.fi}
}

% \begingroup
% \renewcommand\thefootnote{}
% \footnotetext{\small $^{*}$ Corresponding author}
% \endgroup

%\maketitle
%
%%%% list of authors for the TOC (use if author list has to be modified)
%\tocauthor{Ivar Ekeland, Roger Temam, Jeffrey Dean, David Grove,Craig Chambers, Kim B. Bruce, and Elisa Bertino}
%\institute{University of Turku, Finland}\\
%\email{author.name1@gmail.com}
%\and
%University Name, City, Country\\
%\email{author.name1@gmail.com}}

\maketitle              % typeset the title of the contribution

\begin{abstract}
Chain-of-Thought (CoT) prompting has been used to enhance the reasoning capability of LLMs. However, its reliability in security-sensitive analytical tasks remains insufficiently examined, particularly under structured human evaluation. Alternative approaches, such as model scaling and fine-tuning can be used to help improve performance. These methods are also often costly, computationally intensive, or difficult to audit. In contrast, prompt engineering provides a lightweight, transparent, and controllable mechanism for guiding LLM reasoning.
This study proposes a structured prompt engineering framework designed to strengthen CoT reasoning integrity while improving security threat and attack detection reliability in local LLM deployments. The framework includes 16 factors grouped into four core dimensions: (1) Context and Scope Control, (2) Evidence Grounding and Traceability, (3) Reasoning Structure and Cognitive Control, and (4) Security-Specific Analytical Constraints. Rather than optimizing the wording of the prompt heuristically, the framework introduces explicit reasoning controls to mitigate hallucination and prevent reasoning drift, as well as strengthening interpretability in security-sensitive contexts.
Using DDoS attack detection in SDN traffic as a case study, multiple model families were evaluated under structured and unstructured prompting conditions. Pareto frontier analysis and ablation experiments demonstrate consistent reasoning improvements (up to 40\% in smaller models) and stable accuracy gains across scales. Human evaluation with strong inter-rater agreement (Cohen’s $\kappa > 0.80$) confirms robustness. The results establish structured prompting as an effective and practical approach for reliable and explainable AI-driven cybersecurity analysis.

\end{abstract}
\section{Introduction}

Large language models (LLMs) have become vital tools in supporting cybersecurity practices, from threat detection, incident analysis, to automated investigation \cite{shenoy2024extended,zhang2025llms,atlam2025llms,ferrag2024generative}. However, compared to traditional rule-based and machine learning approaches, LLMs employ natural language reasoning to interpret and generate analytical conclusions, with the quality of their reasoning being significantly influenced by the prompt content. Within high-risk cybersecurity scenarios, LLMs might result in inaccurate assessments, overreactions, or missing of key evidence because without a clear structure or governance, there is a possibility of logical leaps, redundant reasoning, and even hallucinations \cite{kasri2025vulnerability,sood2025hallucinations}. Therefore, how to efficiently supervise LLMs to provide reliable and robust reasoning that is consistent with the security analysis workflow has emerged as a significant research challenge.

Although significant research has been done on prompt engineering, research has mostly focused on reasoning formats and algorithmic optimisation, with little attention paid to constraining the reasoning process and aligning outputs with the cognitive logic of human analysts \cite{shenoy2024extended,huang2024prompt}. Security analysts working in cybersecurity tasks demand not only model outputs, but also transparent, traceable, evidence-based reasoning processes to support in verification and accountability \cite{ahi2025llms}. As an outcome, the capacity to explain reasoning steps via a Chain-of-Thought (CoT) is especially important, as it additionally serves as an explanatory foundation for the LLMs' results. The reasoning steps offer structural support for auditing the validity and consistency of the conclusions. Evidently, prompt design influences not just reasoning accuracy, but also when LLMs could meaningfully assist security analysts in an understandable and accountable approach \cite{wei2022chain,habibzadeh2025soc}.

Deep research on how various prompting strategies affect human-centered reasoning in cybersecurity applications is still limited. Due to the lack of control over prompt components in existing approaches, outcomes are often unstable or difficult to reproduce. These limitations hinder our ability to evaluate the effectiveness of various prompting strategies and constrain the reliable application of LLMs in specialised security scenarios.

Our case study is based on the DDoS SDN dataset to formulate and evaluate the structured CoT prompt engineering framework, which enhances reasoning performance and integrity across various sizes of local LLMs for cybersecurity applications. We identify the following research questions for this study:

\begin{itemize}
    \item[\textbf{RQ1:}] Does structured prompt engineering improve reasoning quality and detection accuracy in local LLMs compared to unstructured prompting?
    
    \item[\textbf{RQ2:}] How does structured prompting influence the trade-off between detection accuracy and reasoning interpretability across different model sizes?
\end{itemize}

This study makes three main contributions. First, we propose a structured prompt engineering framework consisting of 16 control factors grouped into four reasoning domains—Context and Scope Control, Evidence Grounding and Traceability, Reasoning Structure and Cognitive Control, and Security-Specific Analytical Constraints, to systematically guide reasoning in local LLM execution. Second, through empirical evaluation on DDoS attack detection in SDN traffic, we demonstrate that structured prompting significantly improves reasoning quality while maintaining stable detection accuracy across multiple model families and scales. Finally, we introduce a human-centred reasoning evaluation methodology with strong inter-rater reliability, ensuring that reasoning improvements are measurable, consistent, and reproducible. Together, these contributions position structured prompt engineering as a practical and reliable approach for trustworthy AI-driven cybersecurity analysis.

The remainder of this paper is organized as follows: Section~\ref{sec:related} reviews related work; Section~\ref{sec:framework} presents the structured prompt framework; Section~\ref{sec:experiment} describes the experimental methodology; Section~\ref{sec:evaluation} reports the data analysis and research results. Section~\ref{sec:discussion} discusses the findings. Finally, Section~\ref{sec:conclusion} concludes the paper.

\section{Related Works}
\label{sec:related}

\textbf{LLMs in Security-Critical Domains.} 
Large Language Models (LLMs) are commonly deployed as either cloud-based proprietary systems or locally hosted open-source models. Cloud-based LLMs typically run on large-scale computational infrastructure, providing stronger inference capabilities and faster update cycles, and tend to be therefore widely employed for compute-intensive inference tasks \cite{zhao2023survey}. Although, relying on external platforms brings significant risks including privacy breaches, data leaking, and compliance \cite{chen2024combating,brown2022privacy,balloccu2024leak,adeseye2025local}. In contrast, an increasing amount of research evaluates the viability of locally hosted LLMs. For cases with high data sensitivity, such as cybersecurity, local deployment ensures that data processing is wholly internal to the organization, ensuring compliance with regulations for privacy protection and data control \cite{balloccu2024leak,adeseye2025local,montagna2023data,kumar2024local}. However, locally hosted LLMs have limitations because they are commonly restricted in scale. Locally hosted LLMs typically lag behind cloud-based LLMs in terms of parameter scale and alignment, and the reasoning process is more vulnerable to the effect of input methods \cite{mohsin2025limits,askell2021assistant,wolf2023alignment}. Following this, the expression of prompts has a big impact on the reliability and consistency of the locally hosted LLMs' reasoning \cite{wang2024prompt,barrie2024stability}. In conclusion, our study mainly focuses on discussing the impact of prompt structure on the reasoning quality of locally hosted LLMs, and aims to propose an inference control strategy which works better for security tasks based on this.

\medskip
\noindent
\textbf{Prompt Engineering in Cybersecurity.}
Prompt engineering is a crucial research direction for LLMs in cybersecurity. Existing research has shown that constructed prompts could improve LLM performance in tasks such as traffic analysis, anomaly detection, and forensic investigation, thus better meeting the needs of cybersecurity analysis (such as security automation, threat assessment, and incident response) \cite{shenoy2024extended,huang2024prompt,meda2025integrating}. Among common approaches, Chain-of-Thought (CoT) and In-Context Learning (ICL) are widely used. CoT preserves logical coherence by displaying the reasoning process, whereas ICL uses examples to assist the LLM adapt to new tasks \cite{shenoy2024extended,jia2025comprehensive,wang2025trust}.
Meanwhile, some studies explore the use of more standardized prompt templates in specific cybersecurity scenarios to reduce ambiguity in the generation of inference chains by LLMs and make reasoning steps easier to reproduce. This method helps improve the auditability of output and make it easier to integrate into existing workflows \cite{priescu2025prompt,iyenghar2025feasibility}. However, existing work focuses more on performance improvement and less on the control of the inference process itself. Although the application of prompt engineering in security tasks has shown prospects, there is still lack of in-depth discussions on the traceability of reasoning, the reliability of evidence citation, and the degree to which different prompt structures match cybersecurity analysis needs. To address this issue, our study provides a multidimensional prompt design strategy for real-world security scenarios, aiming to enhance the reliability and analytical accuracy of the locally hosted LLMs in complex tasks.

\medskip
\noindent
\textbf{Human-Centric LLM Reasoning.}
LLMs often produce redundancy, over-thinking, or inference bias when lacking clear task guidance \cite{mohsin2025limits,taveekitworachai2024nullshot,singh2025batch,ahn2025reverse,adeseye2026hallucination}. According to some research, introducing structured reasoning paradigms for LLMs can improve the model's reasoning quality without spending additional training expenses. For example, some research build on how humans manage complex tasks, emphasising the importance of initially developing an overall cognitive framework to confine subsequent reasoning processes, hence reducing the impact of irrelevant information on the analytical process \cite{xu2025reason,plaat2025multi}. Other studies focus on the transparency of the reasoning process, arguing that clear intermediate steps help improve the consistency and testability of model reasoning \cite{zeng2023consistent,lee2025reasoning}. In addition to these difficulties, cybersecurity scenarios place higher demands on the reasoning process itself. Artificial intelligence model analysis results often require a clear chain of evidence and allocation of responsibility. Vague or unverifiable reasoning paths could reduce the credibility of judgments and increase false positives or auditing difficulties \cite{osholake2024human,mariam2024human,panteli2025being}. Therefore, the reasoning process that can be traced and verified is of particular importance in security analysis. Overall, we believe that from a human-centered perspective, reasoning integrity requires three properties: (1) procedural transparency, (2) evidence alignment, and (3) consistency with established analytical workflows. These requirements extend beyond raw accuracy metrics and emphasize auditability and explainability in operational settings. In our research, we have taken these factors into account when designing our multi-dimensional prompt strategy; at the same time, we have also incorporated these aspects into the Human-Evaluated Reasoning Quality metric when evaluating the reasoning performance of LLMs.

\medskip
\noindent
\textbf{Chain-of-Thought (CoT) Reasoning.}
The Chain-of-Thought (CoT) is a prompt engineering technique for LLMs. By providing clear intermediate reasoning steps in natural language, CoT can separate complicated problems into multi-step reasoning processes, which enhances the LLM's analytical capabilities and interpretability \cite{wei2022chain,wang2025trust}. As CoT is more often used in complex reasoning tasks, more and more studies are beginning to focus on its construction methods, which is influencing factors, enhancement strategies, and performance in practical applications \cite{jia2025comprehensive,wang2025trust}. At the methodology level, Q Chen et al. categorized deep reasoning into three types based on reasoning expression forms: natural language reasoning, structured language reasoning, and latent space reasoning. These classifications provide a crucial foundation for understanding the developmental trajectory of multi-step reasoning approaches \cite{chen2025towards}. However, the application of the regulatory effects of different constraint mechanisms on reasoning behavior remains lacking in cybersecurity tasks. Next, while CoT increases the depth of reasoning, unconstrained CoT may also amplify intermediate steps in illusions or introduce logically coherent but fact-lacking reasoning chains \cite{adeseye2026hallucination,cheng2025cot}. Therefore, we believe that constrained CoT design is meaningful in security-sensitive applications.

\begin{figure}[H]
    \centering
    \includegraphics[width=0.69\linewidth]{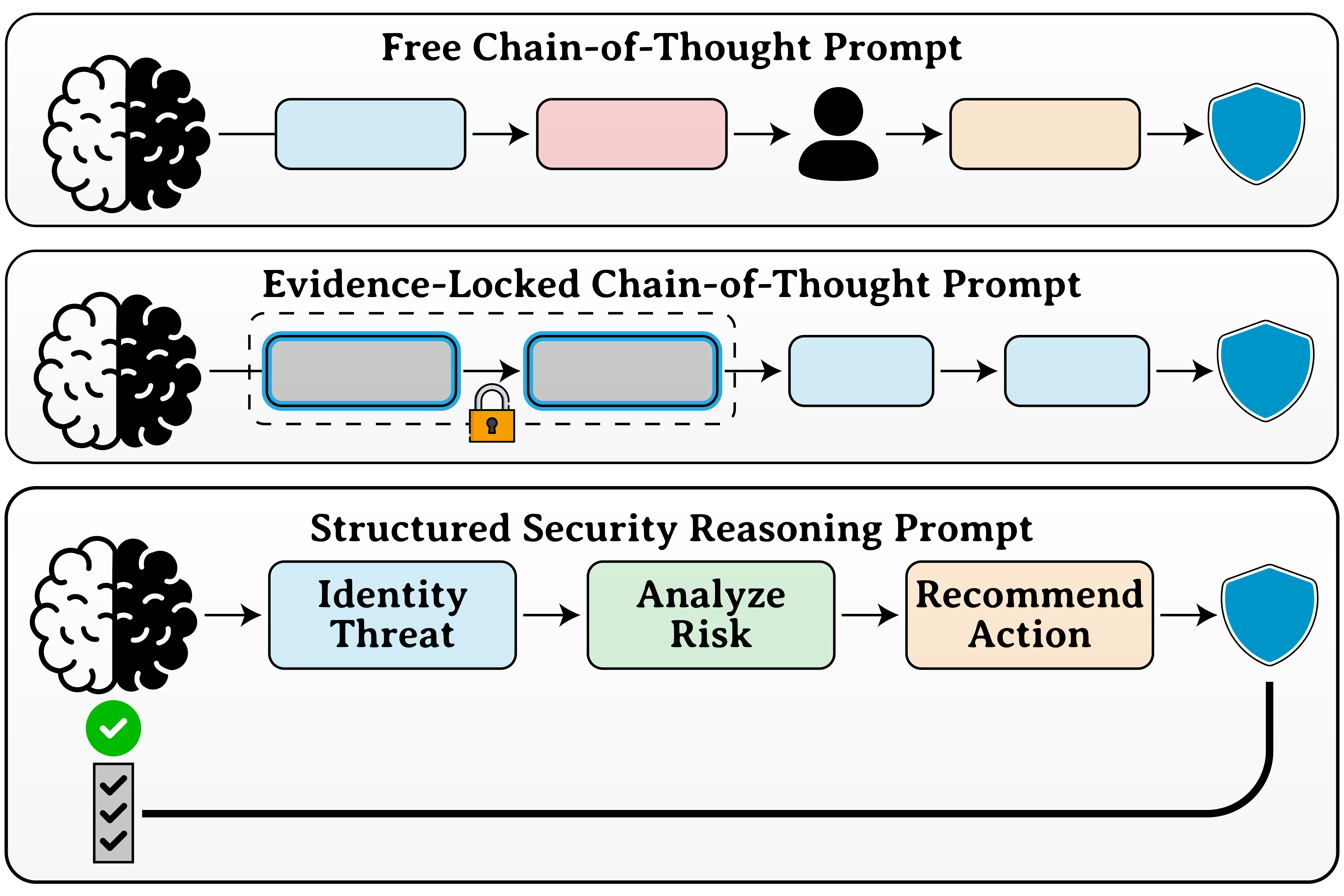}
    \caption{
    \textbf{Three types of CoT prompts:}
    \textbf{(1)Free CoT Prompt}: The step-by-step reasoning approach with minimal structural constraints emphasizes flexibility;
    \textbf{(2)Evidence-Locked CoT Prompt}: Incorporating consistency constraints into the reasoning process, requiring conclusions to be based on given information, increasing credibility and reducing hallucinations;
    \textbf{(3)Structured Security Reasoning Prompt}: The reasoning process follows the actual workflow of security analysis (e.g., threat detection → risk analysis → action recommendation) to ensure the model output is consistent with the decision-making logic of human security experts.  
    }
    \label{fig:typesofCOT}
\end{figure}

\medskip
\noindent
\textbf{Prompting Strategies Design.}
In terms of prompt strategy design, we divided it onto two sections: input structure and output sensitivity control. Although zero-shot prompting, structured prompting, role-play prompting, and other techniques have been extensively explored, there still has been limited evaluation of the reasoning quality and practical applicability in cybersecurity practice \cite{kojima2022large,kong2024better,li2025structured}. Second, human-centered demands such as the interpretability and traceability of evidence receive insufficient attention. With these considerations, our study views prompt design as a structured control mechanism used to constrain the reasoning process and maintain logical integrity in locally hosted LLMs. We propose three representative strategies (see Fig. \ref{fig:typesofCOT}), covering a continuous range from free reasoning to evidence-constrained reasoning to fully structured reasoning in cybersecurity tasks. This enables us to systematically compare the impact of different constraint strengths on the reliability, auditability, and operational applicability of reasoning. On the input design of our research, we built three different dimensions of structured prompt frameworks to adapt to various task needs; in the experimental evaluation phase, we concentrated on analysing the performance of Structured Security Reasoning Prompt.

The output of LLM is highly sensitive to prompt design, including the system-level instruction and the user-level task description. The system prompt defines global behavioral constraints, tone, and role alignment, whereas the user prompt determines task-specific reasoning scope \cite{neumann2025position,adeseye2025prompt}. Improperly constrained prompts may lead to over-reasoning, hallucinations, or logically inconsistent outputs. In cybersecurity workflows, this sensitivity has direct operational implications, such as potentially leading to inflated threat assessments, unsupported attribution claims, or false positives \cite{taveekitworachai2024nullshot,singh2025batch,ahn2025reverse}. Therefore, we focus on controlling prompt sensitivity and employing structured prompts to narrow the inference boundary at the output design level. Unlike studies which exclusively analyse answer accuracy, our work explores how prompting strategies and the combinations of them affect the traceability and accountability of evidence in AI-assisted security analysis. As a result, our study extensively evaluates both design strategies.

\section{Structured Prompt Framework Extension}
\label{sec:framework}

\noindent
In cloud-based LLM systems, system-level instructions (e.g., safety alignment, role enforcement, and reasoning control) are typically integrated into the backend architecture. Consequently, the separation between system prompts and user prompts is implicitly managed by the service provider. However, in local LLM deployments, this separation does not exist automatically. The developer must explicitly define global behavioral constraints (system prompt) and task-specific analytical instructions (user prompt). Without such structured separation, local models may over-generalize, hallucinate, ignore scope boundaries, or produce ungrounded reasoning.

To address this limitation, the proposed Prompt Engineering (PE) Framework introduces a structured division between system-level controls and user-level analytical constraints to ensure reliability. Importantly, this framework is not designed merely to optimize prompt accuracy, but to improve the quality of both detection and reasoning. In cybersecurity contexts, detection alone is insufficient; the reasoning behind a detection must be transparent, grounded in dataset features, aligned with known attack taxonomies, and explicitly validated.

\begin{figure}[H]
    \centering
    \includegraphics[width=1\linewidth]{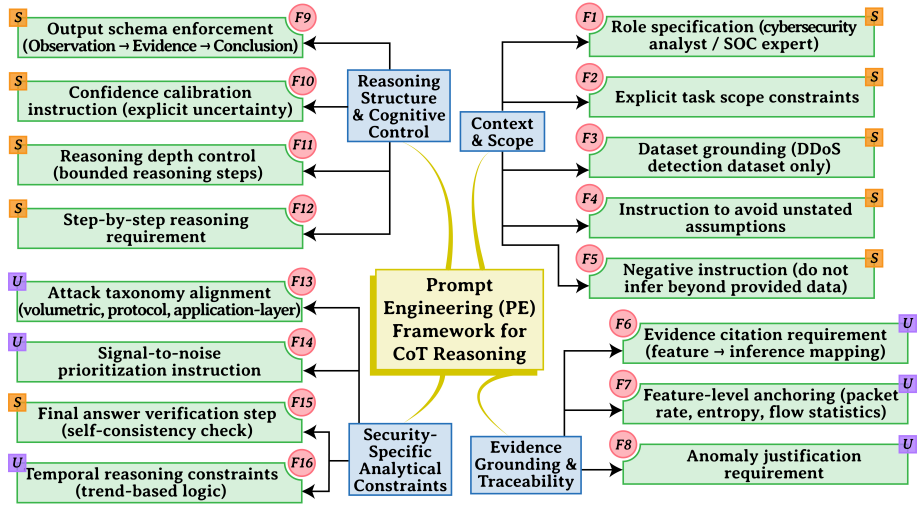}
    \caption{
        \textbf{The Prompt Engineering Framework for CoT Reasoning:}
        \textbf{} organizes 16 reasoning-control factors (F1–F16) across 4 dimensions to reduce hallucination, prevent reasoning drift, and strengthen interpretability in security-sensitive tasks;
        \textbf{Output Control:} System-level factors (S) guide global reasoning, structure, uncertainty calibration, and verification; User-level factors (U) ensure feature grounding, taxonomy alignment, and analytical discipline. 
        }
        
    \label{fig:pe_framework}
\end{figure}

As illustrated in Figure \ref{fig:pe_framework}, the framework integrates sixteen factors (F1–F16), synthesized from literature on chain-of-thought prompting, hallucination mitigation, explainable AI, security analytics, reasoning calibration, and structured decision modeling. The figure visually distinguishes factors assigned to the system prompt (S) and user prompt (U). System-level factors regulate global reasoning behavior, output structure, uncertainty calibration, and validation controls. User-level factors enforce feature grounding, taxonomy alignment, and analytical discipline.

Table \ref{tab:structured_prompt_factors} summarizes these factors, their prompt placement, functional purpose, and supporting literature. Validation is embedded both during reasoning (via evidence citation and feature anchoring) and at the final answer stage (via self-consistency verification), ensuring that both classification and explanation remain trustworthy. Although motivated by local LLM deployment challenges, the framework is model-agnostic and applicable to both local and cloud-based systems.

\begin{table}[H]
\centering

\scriptsize
\caption{Structured Prompt Framework Factors and Prompt Placement}
\label{tab:structured_prompt_factors}
\renewcommand{\arraystretch}{1.2}
\setlength{\tabcolsep}{2.5pt}
\resizebox{\linewidth}{!}{
\begin{tabular}{L{0.75cm} L{3.0cm} L{0.55cm} L{3.4cm} L{3.3cm}}
\hline
\textbf{ID} & \textbf{Factor} & \textbf{P} & \textbf{Purpose} & \textbf{Literature Grounding} \\
\hline
F1  & Role specification (cybersecurity analyst / SOC expert) & S & Defines expert reasoning perspective & Role prompting literature \\
F2  & Explicit task scope constraints & S & Prevents task drift beyond dataset & Prompt constraint research \\
F3  & Dataset grounding (DDoS dataset only) & S & Restricts reasoning to provided data & Hallucination mitigation \\
F4  & Avoid unstated assumptions & S & Reduces unsupported inference & Faithfulness studies \\
F5  & Negative instruction (no inference beyond data) & S & Blocks external knowledge leakage & Prompt safety research \\
\hline
F6  & Evidence citation requirement & U & Enforces feature $\rightarrow$ inference mapping & Explainable AI (XAI) \\
F7  & Feature-level anchoring & U & Grounds reasoning in measurable signals & Security analytics research \\
F8  & Anomaly justification requirement & U & Requires explicit anomaly explanation & Detection theory \\
\hline
F9  & Output schema enforcement (Obs $\rightarrow$ Ev $\rightarrow$ Concl) & S & Standardizes reasoning structure & Structured prompting research \\
F10 & Confidence calibration instruction & S & Encourages uncertainty acknowledgment & Calibration research \\
F11 & Reasoning depth control & S & Prevents overextended reasoning chains & CoT stability literature \\
F12 & Step-by-step reasoning requirement & S & Ensures transparent logical progression & Chain-of-thought literature \\
\hline
F13 & Attack taxonomy alignment & U & Maps reasoning to volumetric / protocol / application layers & Security taxonomy studies \\
F14 & Signal-to-noise prioritization & U & Focuses reasoning on relevant features & Analytical filtering literature \\
F15 & Final answer verification (self-consistency check) & S & Validates final output reliability & Self-consistency research \\
F16 & Temporal reasoning constraints & U & Enforces trend-based anomaly logic & Time-series anomaly detection \\
\hline
\end{tabular}
}
\vspace{-4pt}
\begin{flushleft}
\scriptsize
Note: ID = Framework factor number; S = System prompt; U = User prompt.
\end{flushleft}

\end{table}

\FloatBarrier

\section{Experiment Setting}
\label{sec:experiment}

This study evaluates how different Large Language Models (LLMs) detect and explain cyber attacks using a structured prompt framework. We used a publicly available dataset that clearly labels each sample as either “attack” or “not attack.” This dataset acts as a gold standard, meaning it already contains the correct answers. Using a public dataset ensures transparency, allows other researchers to repeat the study, and increases the credibility of our results. The gold standard labels provide a baseline, which helps us measure how accurate each model is and determine whether our framework improves performance.

\begin{figure}[H]
    \centering
    \includegraphics[width=0.9\linewidth]{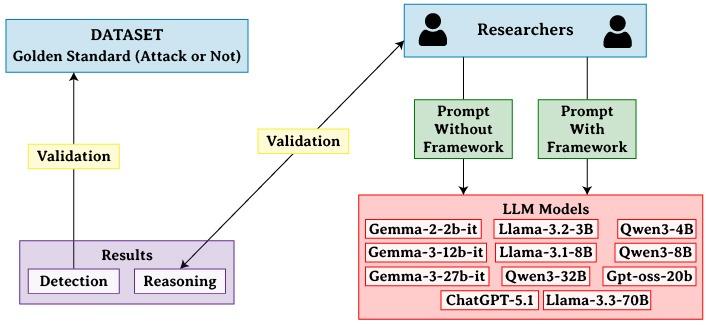}
    \caption{
        \textbf{Experimental Methodology and Prompting Workflow.} 
        \textbf{} This figure shows the experimental workflow. The accuracy of the detection was directly compared with the dataset, while the reasoning part was manually verified by two researchers under different prompt strategies.
        }
    \label{fig:method}
\end{figure}

In this chapter, we provide a detailed introduction  our experiment, please refer to Figure \ref{fig:method}. We tested multiple LLMs from different providers and of different sizes to ensure that the results are not limited to one specific model. Each model was evaluated under two main conditions: using an unstructured prompt (without the framework) and using our structured framework. The prompts were created in two ways. First, we manually designed both unstructured and structured prompts based on our research objectives. Second, we also used ChatGPT to generate both unstructured and structured versions of the prompts. This approach allows us to compare not only the impact of structure versus no structure, but also the difference between human-designed prompts and AI-generated prompts. By doing this, we can clearly measure performance improvements and better understand how prompt design influences the results. 

Two independent researchers were involved in evaluating the outputs and also generating the prompts. Having two researchers reduces bias, improves reliability, and ensures consistent judgment, especially when assessing the reasoning quality of the models. The evaluation focused on both detection accuracy (correct classification) and reasoning quality (clear explanation). This approach ensures that the models are not only correct but also logically sound in their decisions.

\subsection{Dataset}
\looseness=-1
Our research utilized the open-source DDoS (Distributed Denial of Service) SDN dataset \cite{kazin2021ddos} from Kaggle as the experimental data. The original dataset contains multi-dimensional traffic statistics and detailed feature information collected from software-defined networking (SDN) environments. From this dataset, we extracted 400 rows of samples, each consisting of 23 columns, including 3 categorical variables and 20 numerical features. The dataset includes a target variable named “label” to distinguish normal traffic (0) from malicious DDoS attack traffic (1). We selected this dataset because it is publicly available, widely used in cybersecurity research, and provides clearly labeled ground truth data. This improves transparency, reproducibility, and fairness in evaluation. Since the dataset reflects realistic network traffic patterns and attack behavior, it is suitable for testing whether LLMs can correctly interpret structured security data and identify threats. The selection of 400 records ensures a balanced and manageable sample size for systematic experimentation. It is large enough to provide meaningful statistical comparison across different prompting strategies, while still being practical for detailed reasoning evaluation by researchers. This size also helps reduce random bias while allowing consistent validation of detection accuracy and interpretability across models. 

\subsection{Local LLMs}
\looseness=-1
In this study, we incorporated multiple Large Language Models (LLMs) from different families, including Gemma (2B, 3B, 7B), Llama (3.1–8B, 3.2–3B, 3.3–70B), Qwen (3–4B, 3–8B), GPT-OSS (20B), and ChatGPT-5.1. These models were selected to represent diverse architectures, training philosophies, and parameter scales. Using different model families allows us to examine whether our framework performs consistently across varying design strategies rather than being optimized for a single ecosystem. The “B” parameter refers to billions of parameters, which indicates model size and capacity. Smaller models (2B–4B) generally have lower computational requirements but may have limited reasoning depth. Mid-sized models (7B–20B) often provide a balance between efficiency and performance. Large models such as 70B typically demonstrate stronger contextual understanding and advanced reasoning abilities due to their higher representational capacity. By incorporating models across this spectrum, we evaluate how scale influences detection accuracy and interpretability. Our objective is not only to compare raw performance but also to assess how our proposed structured prompting framework improves smaller models and whether larger models inherently require less guidance. This helps us understand scalability, robustness, and the general effectiveness of our framework across lightweight and high-capacity LLM systems.

\subsection{Evaluation Metrics}
The evaluation framework is designed to assess both the reasoning accuracy and the reasoning quality of large language models when performing DDoS attack detection with CoT reasoning. Our framework applies a two-layer evaluation approach, including automatic classification metrics and human-evaluated reasoning metrics. The ground-truth labels provided in the dataset were used as the standard for all classification evaluations. Simultaneously, based on the datasets feature definitions and established DDoS attack features, a structured reasoning evaluation standard was constructed to assess reasoning quality. Two independent reviewers scored the output of each model, and the final result for each reasoning metric was the average of the two reviewers' scores.

\subsubsection{Classification Performance Metrics}

The classification metrics for performance are used to evaluate whether LLMs are capable of correctly identifying between DDoS attacks and normal network traffic \cite{han2025uav}. 

\textbf{Accuracy} represents the proportion of correct predictions out of all samples \cite{tharwat2021classification,vujovic2021metrics}, defined as

\begin{equation}
\text{Accuracy} = \frac{TP + TN}{TP + TN + FP + FN}
\end{equation}

where $TP$ refers to true positive (accurately identified DDoS attack), $TN$ for true negative (correctly classified normal traffic), $FP$ for false positive (normal traffic misidentified as an attack), and $FN$ for false negative (attack missed). This metric measures total detection performance and is useful compared to typical intrusion detection systems (IDS).

\textbf{Precision} represents the proportion of actual attacks out of all instances predicted as attacks \cite{tharwat2021classification,vujovic2021metrics}. Precision is computed as

\begin{equation}
\text{Precision} = \frac{TP}{TP + FP}
\end{equation}

This metric reflects the reliability of attack predictions. In actual security operations, a high false positive rate often increases the workload of analysts, thus affecting overall efficiency; therefore, it is necessary to monitor changes in precision rate.

\textbf{Recall} (known as True response rate or Detection rate) reflects a model's ability to detect real DDoS attacks in a dataset \cite{tharwat2021classification,vujovic2021metrics}. Recall is defined as

\begin{equation}
\text{Recall} = \frac{TP}{TP + FN}
\end{equation}

High Recall is critical for security applications, as undetected attacks may result in service disruption or financial loss.

The \textbf{F1-score} is an evaluation statistic that takes into account both precision and recall \cite{naidu2023metrics,yacouby2020probabilistic}. Its calculation formula is 

\begin{equation}
\text{F1-score} = 2 \times \frac{\text{Precision} \times \text{Recall}}{\text{Precision} + \text{Recall}}
\end{equation}

For cybersecurity cases in which data is unevenly distributed, utilising precision alone might be misleading, which is why it often comes along with the F1 score as measured.

\subsubsection{Reasoning Quality Metrics}

While classification metrics reflect the correctness of results but struggle to reflect the quality, transparency, and reliability of the model's reasoning process. Therefore, we introduce manually evaluated reasoning metrics to evaluate evidence grounding, faithfulness, structural compliance, domain alignment, and confidence calibration \cite{golovneva2022roscoe}. All reasoning metrics are scored ordinally by two independent reviewers using pre-defined scoring criteria, with the final score being the average of the two reviewers' scores.

\textbf{Evidence Grounding Accuracy} evaluates whether the model explicitly anchors its reasoning to observable dataset features such as traffic rates, entropy measures, or flow statistics\cite{golovneva2022roscoe}. \textbf{Reasoning Faithfulness} assesses whether the model introduces unsupported assumptions or hallucinated claims beyond the provided data, addressing risks associated with ungrounded chain-of-thought reasoning \cite{golovneva2022roscoe}. \textbf{Reasoning Structure Compliance} evaluates whether the LLM follows a predefined "observation-evidence-conclusion" pattern, which improves the results' interpretability and enables validation by security analysts. \textbf{Attack Taxonomy Alignment} determines whether the LLM's categorisation results correspond with existing DDoS attack categories and whether key traffic features support them \cite{golovneva2022roscoe}. 

To examine the consistency of human rating, this paper uses the Cohen's Kappa coefficient to measure the degree of consistency among raters, defined as follows:

\begin{equation}
\kappa = \frac{P_o - P_e}{1 - P_e}
\end{equation}

where $P_o$ represents actual consistency between the two raters, and $P_e$ represents incidental consistency. Reporting the Cohen's Kappa coefficient demonstrates that the inferential evaluation results have a certain degree of repeatability and reduces the impact of individual rater bias, thereby improving the reliability of the evaluation process \cite{vieira2010kappa}.

\section{Evaluation}
\label{sec:evaluation}

%\paragraph{Classification Performance Evaluation}
\subsubsection{Classification Performance Evaluation.}
Table \ref{tab:classification_fw} illustrates the reasoning performance of LLMs of varying sizes under two CoT prompting approaches. Overall, ChatGPT Prompt consistently demonstrates superiority over Manual Prompt across all LLMs, contributing marginal improvements across all metrics. These findings indicate that CoT prompts effectively mitigate reasoning bias and enhance model prediction consistency. As LLM parameter size increases, all reasoning performance metrics exhibit a consistent rise, achieving optimal performance on Llama-3.3-70B and ChatGPT-5.1. The results further demonstrate that when local LLama parameters are sufficiently large, its performance approximates or even matches optimal levels, indicating that large-parameter local LLMs retain comparable reasoning capabilities to cloud-based LLMs.

Moreover, the consistent improvement across Precision, Recall, and F1-score demonstrates that CoT prompts significantly enhance the stability of the LLM's end-to-end reasoning process. Increased precision indicates reduced false positives in benign traffic, while improved recall highlights greater sensitivity in detecting DDoS attacks. The concurrent increase in F1-score reflects more balanced and reliable overall predictions. ChatGPT Prompt delivers consistent gains across all three metrics, demonstrating that CoT prompts reduce irrelevant reasoning and potential hallucinations, thereby enhancing the LLM's decision-making quality in security analysis tasks.

\begin{table}[H]
\centering
\scriptsize
\caption{Classification Performance Comparison Across Models and Prompting Strategies (NoFW $\rightarrow$ FW).  Note: P-Prompt, M-Manual and C-ChatGPT.}
\label{tab:classification_fw}
\renewcommand{\arraystretch}{1.2}
\setlength{\tabcolsep}{3pt}
\resizebox{\linewidth}{!}{
\begin{tabular}{llcccc}
\hline
\textbf{Model} & \textbf{P} &
\textbf{Accuracy} &
\textbf{Precision} &
\textbf{Recall} &
\textbf{F1-score} \\
\hline

\multirow{2}{*}{gemma-2B}
 & M  
 & \before{69.8} $\overset{\pctbox{4.0}}{\rightarrow}$ \after{72.6}
 & \before{0.67} $\overset{\pctbox{6.0}}{\rightarrow}$ \after{0.71}
 & \before{0.65} $\overset{\pctbox{12.3}}{\rightarrow}$ \after{0.73}
 & \before{0.66} $\overset{\pctbox{9.1}}{\rightarrow}$ \after{0.72} \\
 & C 
 & \before{71.6} $\overset{\pctbox{4.8}}{\rightarrow}$ \after{75.0}
 & \before{0.69} $\overset{\pctbox{5.8}}{\rightarrow}$ \after{0.73}
 & \before{0.69} $\overset{\pctbox{10.1}}{\rightarrow}$ \after{0.76}
 & \before{0.69} $\overset{\pctbox{7.2}}{\rightarrow}$ \after{0.74} \\

\multirow{2}{*}{Llama-3B}
 & M
 & \before{73.1} $\overset{\pctbox{4.1}}{\rightarrow}$ \after{76.1}
 & \before{0.72} $\overset{\pctbox{4.2}}{\rightarrow}$ \after{0.75}
 & \before{0.69} $\overset{\pctbox{10.1}}{\rightarrow}$ \after{0.76}
 & \before{0.70} $\overset{\pctbox{7.1}}{\rightarrow}$ \after{0.75} \\
& C & \before{75.0} $\overset{\pctbox{4.1}}{\rightarrow}$ \after{78.1}
 & \before{0.74} $\overset{\pctbox{5.4}}{\rightarrow}$ \after{0.78}
 & \before{0.72} $\overset{\pctbox{9.7}}{\rightarrow}$ \after{0.79}
 & \before{0.73} $\overset{\pctbox{6.8}}{\rightarrow}$ \after{0.78} \\

\multirow{2}{*}{Qwen3-4B}
 & M
 & \before{72.6} $\overset{\pctbox{3.9}}{\rightarrow}$ \after{75.4}
 & \before{0.71} $\overset{\pctbox{4.2}}{\rightarrow}$ \after{0.74}
 & \before{0.68} $\overset{\pctbox{10.3}}{\rightarrow}$ \after{0.75}
 & \before{0.69} $\overset{\pctbox{7.2}}{\rightarrow}$ \after{0.74} \\
 & C
 & \before{74.4} $\overset{\pctbox{3.8}}{\rightarrow}$ \after{77.2}
 & \before{0.73} $\overset{\pctbox{4.1}}{\rightarrow}$ \after{0.76}
 & \before{0.71} $\overset{\pctbox{8.5}}{\rightarrow}$ \after{0.77}
 & \before{0.72} $\overset{\pctbox{5.6}}{\rightarrow}$ \after{0.76} \\

\hline

\multirow{2}{*}{gemma-12B}
 & M
 & \before{78.9} $\overset{\pctbox{2.8}}{\rightarrow}$ \after{81.1}
 & \before{0.77} $\overset{\pctbox{2.6}}{\rightarrow}$ \after{0.79}
 & \before{0.74} $\overset{\pctbox{8.1}}{\rightarrow}$ \after{0.80}
 & \before{0.75} $\overset{\pctbox{5.3}}{\rightarrow}$ \after{0.79} \\
 & C
 & \before{80.4} $\overset{\pctbox{2.7}}{\rightarrow}$ \after{82.6}
 & \before{0.79} $\overset{\pctbox{3.8}}{\rightarrow}$ \after{0.82}
 & \before{0.78} $\overset{\pctbox{6.4}}{\rightarrow}$ \after{0.83}
 & \before{0.78} $\overset{\pctbox{5.1}}{\rightarrow}$ \after{0.82} \\

\multirow{2}{*}{Llama-8B}
 & M
 & \before{80.5} $\overset{\pctbox{2.2}}{\rightarrow}$ \after{82.3}
 & \before{0.79} $\overset{\pctbox{2.5}}{\rightarrow}$ \after{0.81}
 & \before{0.76} $\overset{\pctbox{6.6}}{\rightarrow}$ \after{0.81}
 & \before{0.77} $\overset{\pctbox{5.2}}{\rightarrow}$ \after{0.81} \\
 & C
 & \before{82.0} $\overset{\pctbox{2.4}}{\rightarrow}$ \after{84.0}
 & \before{0.81} $\overset{\pctbox{3.7}}{\rightarrow}$ \after{0.84}
 & \before{0.80} $\overset{\pctbox{6.3}}{\rightarrow}$ \after{0.85}
 & \before{0.80} $\overset{\pctbox{5.0}}{\rightarrow}$ \after{0.84} \\

\multirow{2}{*}{Qwen3-8B}
 & M
 & \before{79.8} $\overset{\pctbox{2.8}}{\rightarrow}$ \after{82.0}
 & \before{0.78} $\overset{\pctbox{2.6}}{\rightarrow}$ \after{0.80}
 & \before{0.75} $\overset{\pctbox{8.0}}{\rightarrow}$ \after{0.81}
 & \before{0.76} $\overset{\pctbox{5.3}}{\rightarrow}$ \after{0.80} \\
 & C
 & \before{81.3} $\overset{\pctbox{2.7}}{\rightarrow}$ \after{83.5}
 & \before{0.80} $\overset{\pctbox{3.8}}{\rightarrow}$ \after{0.83}
 & \before{0.79} $\overset{\pctbox{6.3}}{\rightarrow}$ \after{0.84}
 & \before{0.79} $\overset{\pctbox{5.1}}{\rightarrow}$ \after{0.83} \\

\hline

\multirow{2}{*}{gemma-27B}
 & M
 & \before{84.6} $\overset{\pctbox{1.4}}{\rightarrow}$ \after{85.8}
 & \before{0.83} $\overset{\pctbox{2.4}}{\rightarrow}$ \after{0.85}
 & \before{0.81} $\overset{\pctbox{4.9}}{\rightarrow}$ \after{0.85}
 & \before{0.81} $\overset{\pctbox{4.9}}{\rightarrow}$ \after{0.85} \\
 & C
 & \before{86.1} $\overset{\pctbox{1.5}}{\rightarrow}$ \after{87.4}
 & \before{0.85} $\overset{\pctbox{2.4}}{\rightarrow}$ \after{0.87}
 & \before{0.84} $\overset{\pctbox{4.8}}{\rightarrow}$ \after{0.88}
 & \before{0.84} $\overset{\pctbox{3.6}}{\rightarrow}$ \after{0.87} \\

\multirow{2}{*}{Qwen3-32B}
 & M
 & \before{86.2} $\overset{\pctbox{1.6}}{\rightarrow}$ \after{87.6}
 & \before{0.85} $\overset{\pctbox{2.4}}{\rightarrow}$ \after{0.87}
 & \before{0.83} $\overset{\pctbox{4.8}}{\rightarrow}$ \after{0.87}
 & \before{0.83} $\overset{\pctbox{4.8}}{\rightarrow}$ \after{0.87} \\
 & C
 & \before{87.7} $\overset{\pctbox{1.7}}{\rightarrow}$ \after{89.2}
 & \before{0.87} $\overset{\pctbox{2.3}}{\rightarrow}$ \after{0.89}
 & \before{0.86} $\overset{\pctbox{4.7}}{\rightarrow}$ \after{0.90}
 & \before{0.86} $\overset{\pctbox{3.5}}{\rightarrow}$ \after{0.89} \\

\multirow{2}{*}{gpt-20B}
 & M
 & \before{88.9} $\overset{\pctbox{1.3}}{\rightarrow}$ \after{90.1}
 & \before{0.89} $\overset{\pctbox{1.1}}{\rightarrow}$ \after{0.90}
 & \before{0.87} $\overset{\pctbox{2.3}}{\rightarrow}$ \after{0.89}
 & \before{0.87} $\overset{\pctbox{2.3}}{\rightarrow}$ \after{0.89} \\
 & C
 & \before{90.2} $\overset{\pctbox{1.3}}{\rightarrow}$ \after{91.4}
 & \before{0.90} $\overset{\pctbox{2.2}}{\rightarrow}$ \after{0.92}
 & \before{0.88} $\overset{\pctbox{3.4}}{\rightarrow}$ \after{0.91}
 & \before{0.89} $\overset{\pctbox{2.2}}{\rightarrow}$ \after{0.91} \\

\multirow{2}{*}{Llama-70B}
 & M
 & \before{89.5} $\overset{\pctbox{1.6}}{\rightarrow}$ \after{90.9}
 & \before{0.90} $\overset{\pctbox{1.1}}{\rightarrow}$ \after{0.91}
 & \before{0.88} $\overset{\pctbox{2.3}}{\rightarrow}$ \after{0.90}
 & \before{0.88} $\overset{\pctbox{2.3}}{\rightarrow}$ \after{0.90} \\
 & C
 & \before{91.0} $\overset{\pctbox{1.3}}{\rightarrow}$ \after{92.2}
 & \before{0.92} $\overset{\pctbox{1.1}}{\rightarrow}$ \after{0.93}
 & \before{0.90} $\overset{\pctbox{2.2}}{\rightarrow}$ \after{0.92}
 & \before{0.90} $\overset{\pctbox{2.2}}{\rightarrow}$ \after{0.92} \\

\multirow{2}{*}{ChatGPT}
 & M
 & \before{91.0} $\overset{\pctbox{1.1}}{\rightarrow}$ \after{92.0}
 & \before{0.91} $\overset{\pctbox{1.1}}{\rightarrow}$ \after{0.92}
 & \before{0.89} $\overset{\pctbox{2.2}}{\rightarrow}$ \after{0.91}
 & \before{0.89} $\overset{\pctbox{2.2}}{\rightarrow}$ \after{0.91} \\
 & C
 & \before{92.4} $\overset{\pctbox{1.0}}{\rightarrow}$ \after{93.3}
 & \before{0.93} $\overset{\pctbox{1.1}}{\rightarrow}$ \after{0.94}
 & \before{0.91} $\overset{\pctbox{2.2}}{\rightarrow}$ \after{0.93}
 & \before{0.91} $\overset{\pctbox{2.2}}{\rightarrow}$ \after{0.93} \\

\hline
\end{tabular}
}
\vspace{-4pt}
\begin{flushleft}
\scriptsize
Note: NoFW = No Framework; FW = With Framework.
\end{flushleft}

\end{table}

\FloatBarrier

In summary, both CoT prompt design and LLM size impact reasoning performance. However, experimental results show that chain-of-thought prompts deliver consistent and significant performance gains across all LLMs, remaining effective even in larger size LLMs. The results of our research show that CoT prompt engineering is capable of making up for the limitations of small local LLMs and improve their performance in secure reasoning tasks, especially in terms of reliability and stability.

\begin{table}[!t]
\centering
\scriptsize
\caption{Human-Evaluated Reasoning Quality Scores Across Models and Prompting Strategies (NoFW $\rightarrow$ FW). Note: P-Prompt, M-Manual and C-ChatGPT.}
\label{tab:reasoning_fw}
\renewcommand{\arraystretch}{0.6}
\setlength{\tabcolsep}{3pt}
\resizebox{\linewidth}{!}{
\begin{tabular}{llccccc}
\hline
\textbf{Models} & \textbf{P} &
\textbf{Evidence} &
\textbf{Faithfulness} &
\textbf{Structure} &
\textbf{Taxonomy} \\
\hline

\multirow{2}{*}{gemma-2b}
 & M
 & \before{0.72} $\overset{\pctboxL{45.8}}{\rightarrow}$ \after{1.05}
 & \before{0.70} $\overset{\pctboxL{45.7}}{\rightarrow}$ \after{1.02}
 & \before{0.76} $\overset{\pctboxL{47.4}}{\rightarrow}$ \after{1.12}
 & \before{0.68} $\overset{\pctboxL{44.1}}{\rightarrow}$ \after{0.98} \\
 & C
 & \before{0.86} $\overset{\pctboxL{37.2}}{\rightarrow}$ \after{1.18}
 & \before{0.84} $\overset{\pctboxL{36.9}}{\rightarrow}$ \after{1.15}
 & \before{0.91} $\overset{\pctboxL{37.4}}{\rightarrow}$ \after{1.25}
 & \before{0.82} $\overset{\pctboxL{34.1}}{\rightarrow}$ \after{1.10} \\

\multirow{2}{*}{Llama-3B}
 & M
 & \before{0.80} $\overset{\pctboxL{40.0}}{\rightarrow}$ \after{1.12}
 & \before{0.78} $\overset{\pctboxL{38.5}}{\rightarrow}$ \after{1.08}
 & \before{0.84} $\overset{\pctboxL{42.9}}{\rightarrow}$ \after{1.20}
 & \before{0.76} $\overset{\pctboxL{38.2}}{\rightarrow}$ \after{1.05} \\
 & C
 & \before{0.94} $\overset{\pctboxL{34.0}}{\rightarrow}$ \after{1.26}
 & \before{0.92} $\overset{\pctboxL{33.7}}{\rightarrow}$ \after{1.23}
 & \before{0.98} $\overset{\pctboxL{34.7}}{\rightarrow}$ \after{1.32}
 & \before{0.90} $\overset{\pctboxL{31.1}}{\rightarrow}$ \after{1.18} \\

\multirow{2}{*}{Qwen3-4B}
 & M
 & \before{0.78} $\overset{\pctboxL{38.5}}{\rightarrow}$ \after{1.08}
 & \before{0.76} $\overset{\pctboxL{36.8}}{\rightarrow}$ \after{1.04}
 & \before{0.82} $\overset{\pctboxL{43.9}}{\rightarrow}$ \after{1.18}
 & \before{0.74} $\overset{\pctboxL{37.8}}{\rightarrow}$ \after{1.02} \\
 & C
 & \before{0.92} $\overset{\pctboxL{32.6}}{\rightarrow}$ \after{1.22}
 & \before{0.90} $\overset{\pctboxL{31.1}}{\rightarrow}$ \after{1.18}
 & \before{0.96} $\overset{\pctboxL{33.3}}{\rightarrow}$ \after{1.28}
 & \before{0.88} $\overset{\pctboxL{29.5}}{\rightarrow}$ \after{1.14} \\

\hline

\multirow{2}{*}{gemma-12b}
 & M
 & \before{1.02} $\overset{\pctboxL{27.5}}{\rightarrow}$ \after{1.30}
 & \before{1.00} $\overset{\pctboxL{26.0}}{\rightarrow}$ \after{1.26}
 & \before{1.08} $\overset{\pctboxL{27.8}}{\rightarrow}$ \after{1.38}
 & \before{0.98} $\overset{\pctboxL{24.5}}{\rightarrow}$ \after{1.22} \\
 & C
 & \before{1.21} $\overset{\pctboxL{19.0}}{\rightarrow}$ \after{1.44}
 & \before{1.18} $\overset{\pctboxL{20.3}}{\rightarrow}$ \after{1.42}
 & \before{1.26} $\overset{\pctboxL{19.0}}{\rightarrow}$ \after{1.50}
 & \before{1.16} $\overset{\pctboxL{17.2}}{\rightarrow}$ \after{1.36} \\

\multirow{2}{*}{Llama-8B}
 & M
 & \before{1.09} $\overset{\pctboxL{21.1}}{\rightarrow}$ \after{1.32}
 & \before{1.06} $\overset{\pctboxL{20.8}}{\rightarrow}$ \after{1.28}
 & \before{1.14} $\overset{\pctboxL{22.8}}{\rightarrow}$ \after{1.40}
 & \before{1.04} $\overset{\pctboxL{21.2}}{\rightarrow}$ \after{1.26} \\
 & C
 & \before{1.28} $\overset{\pctboxL{17.2}}{\rightarrow}$ \after{1.50}
 & \before{1.25} $\overset{\pctboxL{18.4}}{\rightarrow}$ \after{1.48}
 & \before{1.32} $\overset{\pctboxL{19.7}}{\rightarrow}$ \after{1.58}
 & \before{1.22} $\overset{\pctboxL{18.0}}{\rightarrow}$ \after{1.44} \\

\multirow{2}{*}{Qwen3-8B}
 & M
 & \before{1.06} $\overset{\pctboxL{26.4}}{\rightarrow}$ \after{1.34}
 & \before{1.04} $\overset{\pctboxL{25.0}}{\rightarrow}$ \after{1.30}
 & \before{1.12} $\overset{\pctboxL{26.8}}{\rightarrow}$ \after{1.42}
 & \before{1.02} $\overset{\pctboxL{25.5}}{\rightarrow}$ \after{1.28} \\
 & C
 & \before{1.25} $\overset{\pctboxL{21.6}}{\rightarrow}$ \after{1.52}
 & \before{1.22} $\overset{\pctboxL{23.0}}{\rightarrow}$ \after{1.50}
 & \before{1.30} $\overset{\pctboxL{23.1}}{\rightarrow}$ \after{1.60}
 & \before{1.20} $\overset{\pctboxL{21.7}}{\rightarrow}$ \after{1.46} \\

\hline

\multirow{2}{*}{gemma-27b}
 & M
 & \before{1.26} $\overset{\pctboxL{12.7}}{\rightarrow}$ \after{1.42}
 & \before{1.24} $\overset{\pctboxL{12.9}}{\rightarrow}$ \after{1.40}
 & \before{1.32} $\overset{\pctboxL{12.1}}{\rightarrow}$ \after{1.48}
 & \before{1.22} $\overset{\pctboxL{13.1}}{\rightarrow}$ \after{1.38} \\
 & C
 & \before{1.44} $\overset{\pctboxL{11.1}}{\rightarrow}$ \after{1.60}
 & \before{1.42} $\overset{\pctboxL{11.3}}{\rightarrow}$ \after{1.58}
 & \before{1.50} $\overset{\pctboxL{10.7}}{\rightarrow}$ \after{1.66}
 & \before{1.40} $\overset{\pctboxL{10.0}}{\rightarrow}$ \after{1.54} \\

\multirow{2}{*}{Qwen3-32B}
 & M
 & \before{1.32} $\overset{\pctboxL{12.1}}{\rightarrow}$ \after{1.48}
 & \before{1.30} $\overset{\pctboxL{12.3}}{\rightarrow}$ \after{1.46}
 & \before{1.38} $\overset{\pctboxL{11.6}}{\rightarrow}$ \after{1.54}
 & \before{1.28} $\overset{\pctboxL{12.5}}{\rightarrow}$ \after{1.44} \\
 & C
 & \before{1.50} $\overset{\pctboxL{9.3}}{\rightarrow}$ \after{1.64}
 & \before{1.48} $\overset{\pctboxL{9.5}}{\rightarrow}$ \after{1.62}
 & \before{1.56} $\overset{\pctboxL{9.0}}{\rightarrow}$ \after{1.70}
 & \before{1.46} $\overset{\pctboxL{8.2}}{\rightarrow}$ \after{1.58} \\

\multirow{2}{*}{gpt-20b}
 & M
 & \before{1.44} $\overset{\pctboxL{9.7}}{\rightarrow}$ \after{1.58}
 & \before{1.42} $\overset{\pctboxL{9.9}}{\rightarrow}$ \after{1.56}
 & \before{1.48} $\overset{\pctboxL{9.5}}{\rightarrow}$ \after{1.62}
 & \before{1.38} $\overset{\pctboxL{10.1}}{\rightarrow}$ \after{1.52} \\
 & C
 & \before{1.60} $\overset{\pctboxL{7.5}}{\rightarrow}$ \after{1.72}
 & \before{1.58} $\overset{\pctboxL{7.6}}{\rightarrow}$ \after{1.70}
 & \before{1.66} $\overset{\pctboxL{7.2}}{\rightarrow}$ \after{1.78}
 & \before{1.56} $\overset{\pctboxL{7.7}}{\rightarrow}$ \after{1.68} \\

\multirow{2}{*}{Llama-70B}
 & M
 & \before{1.50} $\overset{\pctboxL{9.3}}{\rightarrow}$ \after{1.64}
 & \before{1.48} $\overset{\pctboxL{9.5}}{\rightarrow}$ \after{1.62}
 & \before{1.54} $\overset{\pctboxL{9.1}}{\rightarrow}$ \after{1.68}
 & \before{1.44} $\overset{\pctboxL{9.7}}{\rightarrow}$ \after{1.58} \\
 & C
 & \before{1.66} $\overset{\pctboxL{7.2}}{\rightarrow}$ \after{1.78}
 & \before{1.64} $\overset{\pctboxL{7.3}}{\rightarrow}$ \after{1.76}
 & \before{1.72} $\overset{\pctboxL{7.0}}{\rightarrow}$ \after{1.84}
 & \before{1.62} $\overset{\pctboxL{7.4}}{\rightarrow}$ \after{1.74} \\

\multirow{2}{*}{ChatGPT}
 & M
 & \before{1.55} $\overset{\pctboxL{7.1}}{\rightarrow}$ \after{1.66}
 & \before{1.52} $\overset{\pctboxL{7.9}}{\rightarrow}$ \after{1.64}
 & \before{1.58} $\overset{\pctboxL{7.6}}{\rightarrow}$ \after{1.70}
 & \before{1.48} $\overset{\pctboxL{8.1}}{\rightarrow}$ \after{1.60} \\
 & C
 & \before{1.72} $\overset{\pctboxL{5.8}}{\rightarrow}$ \after{1.82}
 & \before{1.70} $\overset{\pctboxL{5.9}}{\rightarrow}$ \after{1.80}
 & \before{1.76} $\overset{\pctboxL{5.7}}{\rightarrow}$ \after{1.86}
 & \before{1.66} $\overset{\pctboxL{7.2}}{\rightarrow}$ \after{1.78} \\

\hline
\end{tabular}
}
\vspace{-4pt}
\begin{flushleft}
\scriptsize
Note: NoFW = No Framework; FW = With Framework.
\end{flushleft}

\end{table}

\FloatBarrier

\subsubsection{Human-Evaluated Reasoning Quality.}
Table \ref{tab:reasoning_fw} is based on 5 human-centric reasoning metrics (Evidence, Faithfulness, Structure, Taxonomy, Confidence), a systemic analysis was conducted on the CoT reasoning outputs of LLMs. Experimental results demonstrate that ChatGPT Prompt performed superiorly to Manual Prompt across all LLMs and all dimensions. These findings indicate that CoT prompts enable LLMs to generate reasoning outputs that surpass the cognitive approaches employed by security analysts.

Chain-of-Thought prompts help LLMs retain the coherence of their reasoning logic during the generating process, improve evidence citation accuracy, and create more clearly structured reasoning chains. This trend is particularly noticeable in smaller-scale local LLMs. Local LLMs (such as Llama-70B) have also approached the performance of some cloud-based models on these human-centric measures, suggesting that with large enough parameter sizes, local models could provide auditable reasoning processes. Overall, structured CoT prompts improve model output stability in terms of understandability, traceability, and consistency, as well as their alignment with the actual requirements of security analysis scenarios.

\section{Discussion}
\label{sec:discussion}
Prompt engineering becomes structurally more important in local LLM deployments because local models do not inherit embedded alignment layers or hidden system-level safeguards that are typically present in commercial cloud systems. In cloud-based models, internal alignment mechanisms can partially compensate for weak prompts. In contrast, local models rely entirely on explicit instructions. This makes them highly prompt-sensitive, particularly in structured analytical tasks such as DDoS attack detection.

In this study, DDoS attack detection using SDN traffic data was employed as the case study. The task involves classifying network traffic as malicious or normal based on numerical flow features. However, in operational security environments, correct classification alone is insufficient. Analysts must understand which traffic characteristics, such as abnormal packet rate, flow irregularities, or anomalous traffic bursts, led to the detection. Therefore, reasoning integrity is operationally critical.

\begin{figure}[H]
    \centering
    \includegraphics[width=0.65\linewidth]{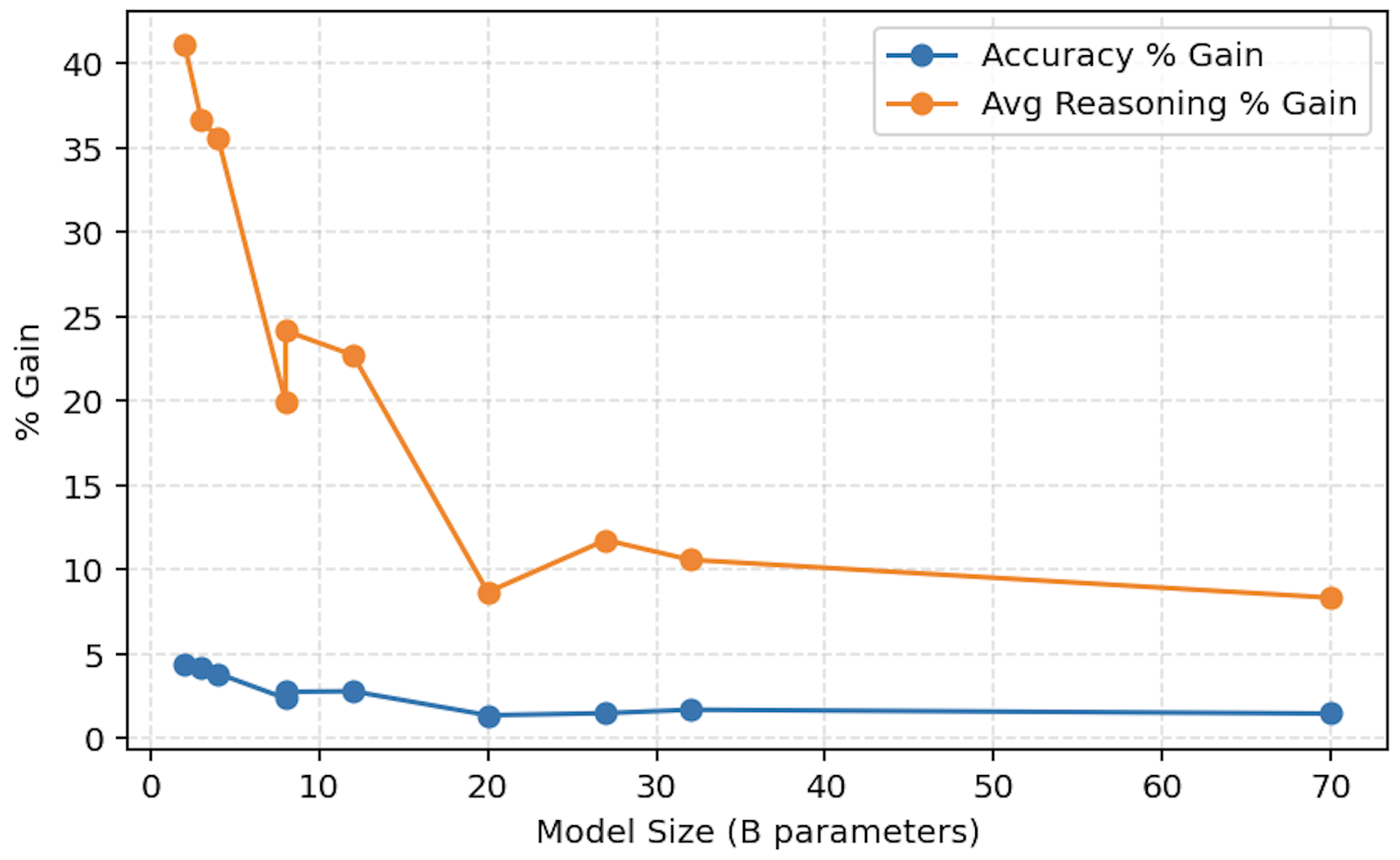}
    \caption{Model Size vs. Relative Performance Gain (Improvement) (FW vs. NoFW) in Detection Accuracy and Reasoning Quality}
    \label{fig:model_size_improvement}
\end{figure}

This distinction is clearly reflected in Figure~\ref{fig:model_size_improvement}. The figure demonstrates that structured prompting improves reasoning quality substantially more than raw detection accuracy. For smaller models (2B–4B), reasoning improvements exceed 35–40\%, while accuracy gains remain below 5\%. Even for larger models (20B–70B), reasoning gains remain between 8–12\%, whereas accuracy improvements are modest (approximately 1–2\%). These results indicate that structured prompting primarily enhances explanation quality rather than classification output. Moreover, the decreasing slope as model size increases suggests diminishing returns: smaller models benefit disproportionately from structured guidance, confirming that prompt sensitivity is higher in low-parameter local models.

\begin{figure}[H]
    \centering
    \includegraphics[width=1\linewidth]{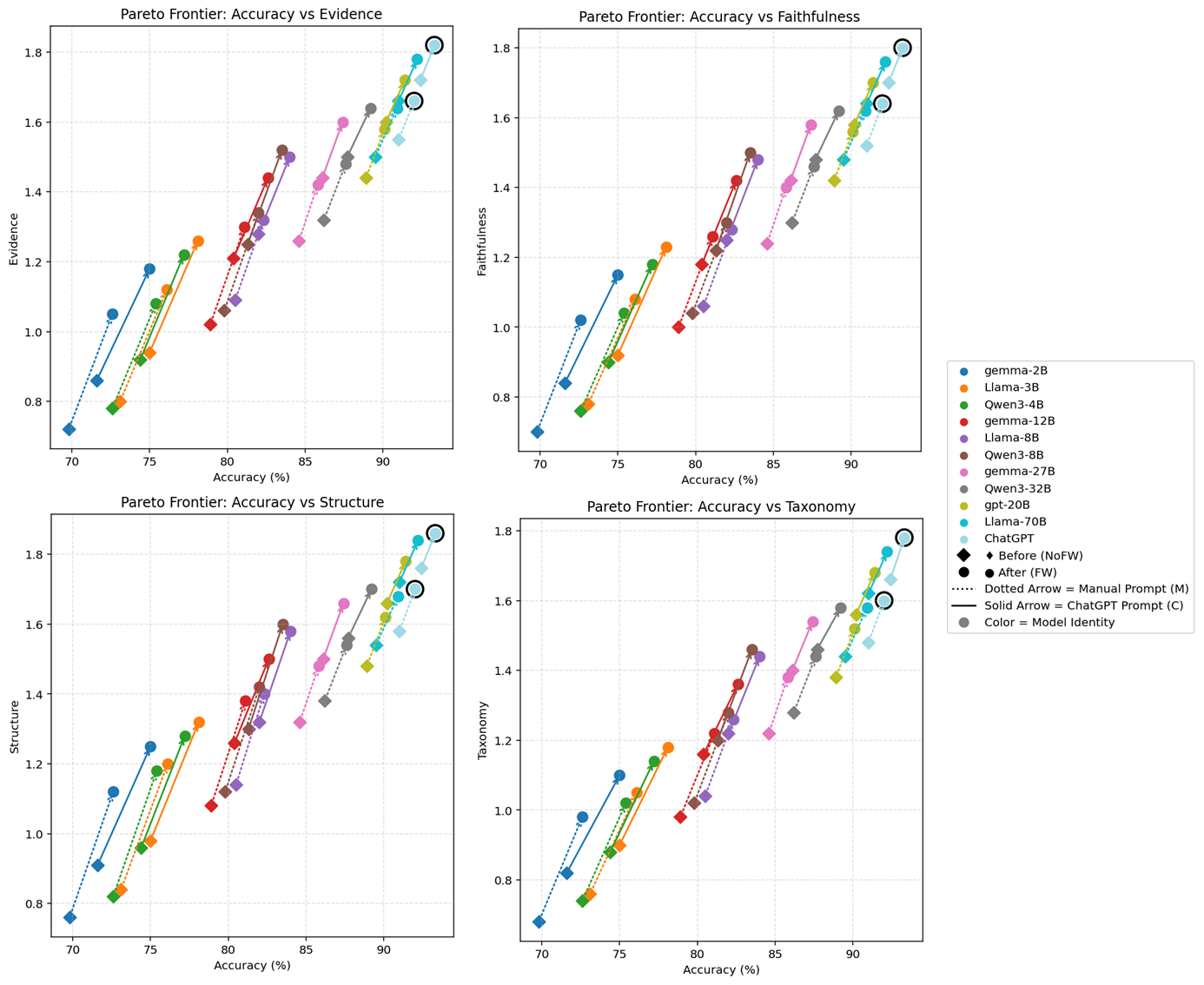}
    \caption{Pareto Frontier Comparison of Security Detection Accuracy and Human-Evaluated Reasoning Dimensions (Evidence, Faithfulness, Structure, Taxonomy) Across Large Language Models with and without the Structured Prompt Framework}
    \label{fig:pareto_frontier}
\end{figure}

Figure~\ref{fig:pareto_frontier} further supports this interpretation through Pareto frontier analysis. Across all four reasoning dimensions—Evidence, Faithfulness, Structure, and Taxonomy—the structured framework consistently shifts model performance toward the upper-right region of the Pareto space. This indicates simultaneous improvement in both detection accuracy and reasoning metrics. Importantly, the observed movement is predominantly vertical (reasoning enhancement) rather than horizontal (accuracy improvement), reinforcing the conclusion that the framework primarily strengthens interpretability. The dotted (manual prompt) and solid (ChatGPT-generated prompt) arrows exhibit similar directional trends, suggesting that the framework’s structural design—not stylistic prompt authoring—drives the performance gains.

The ablation findings align with these visual patterns. Removing evidence-grounding factors resulted in the largest declines in Faithfulness and Evidence scores, confirming that grounding mechanisms are central to maintaining reasoning integrity. Structural controls (e.g., output schema enforcement and step-by-step reasoning) primarily influenced Structure scores, which explains why classification accuracy may remain stable even when explanation quality deteriorates. This reveals a critical operational risk: a model may maintain high detection accuracy while producing weak, unverifiable reasoning.

Additionally, scope constraints and dataset grounding mechanisms prevented reasoning drift. When these factors were removed during ablation, models were more likely to introduce unsupported assumptions, particularly in the interpretation of numerical traffic patterns. This confirms that local LLMs require explicit containment mechanisms to prevent overgeneralization.

The use of both manually designed prompts and ChatGPT-generated prompts strengthens internal validity. Improvements were observed consistently across both prompting strategies, indicating that the structured factors themselves—not stylistic variation—drive the observed gains.

To ensure that reasoning evaluation was reliable, inter-rater agreement was computed using Cohen’s Kappa ($\kappa$). Agreement scores were $\kappa = 0.87$ (Evidence), $\kappa = 0.84$ (Faithfulness), $\kappa = 0.89$ (Structure), and $\kappa = 0.82$ (Taxonomy), indicating strong to very strong agreement. These values confirm that the reasoning improvements illustrated in Figure \ref{fig:pareto_frontier} are reproducible and not dependent on subjective interpretation.

Taken together, the visual evidence and ablation analysis demonstrate that structured prompting does not merely optimize classification accuracy; it fundamentally improves reasoning stability, grounding, and interpretability. In DDoS attack detection, where explanations must justify security decisions, these improvements are operationally more valuable than marginal gains in classification percentage. The proposed framework, therefore, enhances not only performance but also reliability and trustworthiness in local LLM-based cybersecurity systems.

\section{Conclusion}
\label{sec:conclusion}
\looseness=-1
This study proposed a structured prompt engineering framework to improve both detection performance and reasoning quality in local Large Language Models (LLMs). Using DDoS attack detection in SDN traffic as a case study, we demonstrated that classification accuracy alone is insufficient in cybersecurity applications. Transparent, grounded, and verifiable reasoning is operationally essential.

The results show that the proposed framework delivers clear and measurable improvements. Detection accuracy improved consistently across all models (approximately 1–5\%), while reasoning quality showed substantially larger gains, reaching up to 40\% improvement in smaller local models and remaining above 8–12\% even for larger models. Pareto frontier analysis confirmed that models shift toward superior combined accuracy–reasoning trade-offs under the framework, with improvements primarily driven by stronger evidence grounding and reasoning structure. The improvement analysis further revealed that smaller models benefit most from structured prompting, highlighting their higher sensitivity to prompt design and the importance of explicit control mechanisms in local deployments.
The ablation study identified evidence grounding and dataset scope constraints as the most influential factors, confirming that grounding mechanisms are central to preventing hallucination and reasoning drift. Strong inter-rater reliability (Cohen’s $\kappa > 0.80$ across all reasoning dimensions) further validates that the improvements in reasoning quality are consistent and reproducible.
Overall, this research demonstrates that structured prompting is not merely a stylistic enhancement but a necessary methodological control for achieving reliable, explainable, and trustworthy AI-based DDoS attack detection in local LLM systems.

Future work will extend this framework beyond DDoS detection to other cybersecurity tasks such as intrusion detection and malware analysis to validate its generalizability. Further research can explore adaptive prompt structuring based on model size, since smaller local models showed greater sensitivity to structured guidance. Additionally, integrating uncertainty estimation and automated validation mechanisms could further enhance reliability in real-world deployments. Expanding evaluation across more local model families and larger datasets will also strengthen the robustness and scalability of the proposed approach.

\subsubsection*{Acknowledgment.}
This project has received funding from the European Union’s Horizon Europe research and innovation programme under the Marie Skłodowska-Curie grant agreement No 101177564 — HAIF.

% \begin{figure}[H]
%     \centering
%     \includegraphics[width=0.5\linewidth]{acknow.png}
%     \label{fig:acknow}
% \end{figure}

\subsubsection*{Declaration on the Use of Generative AI.}
Language editing and grammar-checking tools were used to improve clarity and readability of the manuscript.

%
% ---- Bibliography ----
%
\clearpage
% \bibliographystyle{splncs03}
% \bibliography{refs}

\end{document}